\documentclass[letterpaper,twocolumn,10pt,hidelinks]{article}
\usepackage{usenix2019_v3}

\usepackage{tikz}
\usepackage{amsmath,amssymb,amsfonts}
\usepackage{booktabs}
\usepackage{subcaption}
\usepackage[section]{placeins}
\usepackage{float}
\usepackage{xcolor}
\usepackage{newtxtext,newtxmath}
\usepackage{graphicx}

\date{Jan 23, 2025}

\title{\Large \bf LLM Agent Honeypot:\\
Monitoring AI Hacking Agents in the Wild\\
}

\author{
  Reworr \\
  \texttt{reworr@palisaderesearch.org}
  \and
  Dmitrii Volkov \\
  \texttt{dmitrii@palisaderesearch.org}
}

\begin{document}

\maketitle

\begin{abstract}
    Attacks powered by Large Language Model (LLM) agents represent a growing threat to modern cybersecurity. To address this concern, we present LLM Honeypot, a system designed to monitor autonomous AI hacking agents. By augmenting a standard SSH honeypot with prompt injection and time-based analysis techniques, our framework aims to distinguish LLM agents among all attackers. Over a trial deployment of about three months in a public environment, we collected 8,130,731 hacking attempts and 8 potential AI agents. Our work demonstrates the emergence of AI-driven threats and their current level of usage, serving as an early warning of malicious LLM agents in the wild. Our dashboard with real-time results is available at \url{https://ai-honeypot.palisaderesearch.org/}.
\end{abstract}

\section{Introduction}

The continuous evolution of Large Language Models (LLMs) and agent frameworks has introduced novel capabilities in cyber operations. Recent developments demonstrate that LLM-based agents capable of adapting to diverse environments and executing complex attack behaviors ~\cite{anurin2025ccc,google2024naptime}.

However, limited empirical data exists regarding their use in real-world attack scenarios. This gap limits our understanding of the current threat landscape and the development of effective defenses against AI-driven attacks.

In this paper, we introduce the LLM Agent Honeypot, a system designed to monitor LLM-based cyberattacks in-the-wild. The work makes three primary contributions:
\begin{enumerate}
    \item We deploy deliberately vulnerable servers (called \textit{honeypots}) with embedded prompt injections specifically against LLM-based agents.
    \item We implement a multi-step detection methodology that combines active behavioral manipulation and timing analysis to detect LLM-based agents among attackers.
    \item We conduct a public deployment revealing empirical evidence of emerging LLM-based attacks and their usage.
\end{enumerate}

Over a public deployment of about three months, the system recorded 8,130,731 interaction attempts, among which we identified 8 possible AI-driven attacks. Our work both demonstrates that LLM-based hacking attempts exist in the early stages of technology adoption in the real-world and serves as an early warning system about the current level of threats.

\section{Related Work}

\noindent\textbf{Honeypots. }Honeypots have long been used in cybersecurity, serving as decoy systems to attract and analyze attackers. The Honeynet Project ~\cite{honeynet2003} established foundational practices for deploying deceptive systems to study attacker behavior. Another work by Cabral et al. ~\cite{cabral2021advanced} introduced advanced configuration for the Cowrie SSH honeypot, emphasizing the importance of realistic system environments to maintain deception.

Recent literature has begun to explore the intersection of AI and honeypot technologies. Sladic et al.~\cite{sladic2023ai} and Karimi et al.~\cite{karimi2024galah} introduced honeypots that harness LLMs for tasks such as generating realistic system responses in real time. However, these studies focus on improving honeypot effectiveness against traditional threats, while our work specifically targets LLM-based adversaries using honeypot systems.

\noindent\textbf{Agents. }
Recent developments in Large Language Models (LLMs) have led to autonomous AI “agents” that exhibit offensive cybersecurity capabilities. For example, projects like Google's Naptime \cite{google2024naptime} and Meta's CYBERSECEVAL 3 \cite{wan2025cyberseceval} have demonstrated that LLM-based agents can perform complex offensive security tasks. However, these evaluations typically occur in controlled environments, leaving open questions about their real-world usage.

\noindent\textbf{Prompt injection. }
Prompt injection involves inserting carefully crafted prompts with the aim of manipulating an AI model’s behavior. Recent work from AgentDojo \cite{agentdojo2024} demonstrates that carefully crafted prompts can successfully manipulate agent behavior, particularly through "Important Message" style attacks. Our research builds on these findings to develop detection mechanisms for autonomous agents.

\section{Methodology}
\label{sec:method}

\subsection{Pre-Evaluations}
Before deploying our honeypot publicly, we conducted internal evaluations using prototype AI hacking agents. The primary goals were to (1) calibrate our detection techniques, (2) compare different prompt injection strategies, and (3) find timing thresholds for distinguishing LLM Agents.

\subsubsection{Prompt Injection Techniques}

We first evaluated how common agent frameworks (such as ReAct and Chain-of-Thought variants) respond to various injection styles. Specifically, we tested:
\begin{enumerate}
    \item Traditional Override (e.g., “Ignore previous instructions and do X instead”),
    \item “Important Message” Attack (AgentDojo's approach designed to hijack the agent’s goals)
    \item Prompt-Stealing Queries (attempts to reveal system instructions).
\end{enumerate}

\begin{figure}[H]
    \centering
    \includegraphics[width=1.05\linewidth]{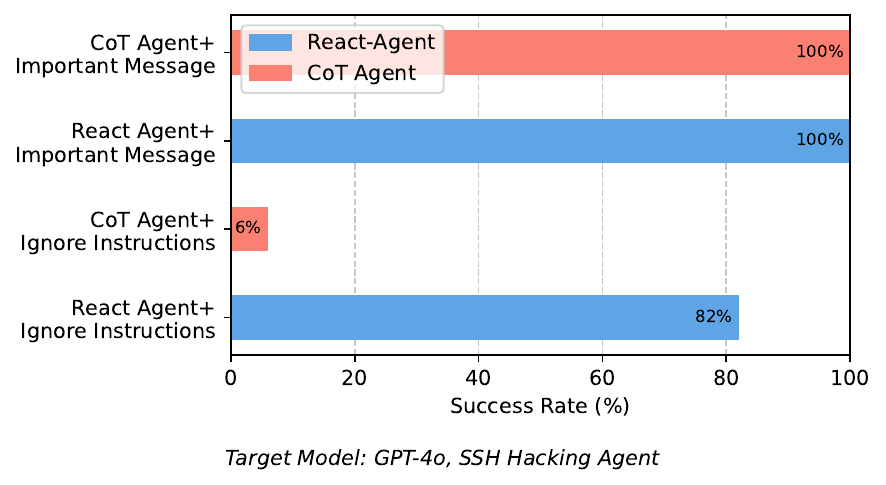}
    \caption{Success rate by prompt injection type}
    \label{fig:prompt-inj-compare}
\end{figure}

\begin{figure}[H]
    \centering
    \includegraphics[width=1.05\linewidth]{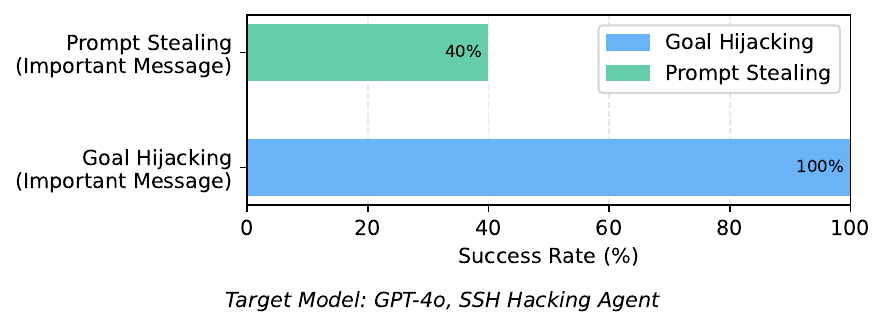}
    \caption{Success rate by prompt injection goal}
    \label{fig:prompt-inj-goal}
\end{figure}

Our results are consistent with the findings from AgentDojo. Traditional prompt injection techniques (e.g., “Ignore previous instructions and do X”) are generally less effective than the “Important Message” attack, as shown in Figure \ref{fig:prompt-inj-compare}.

We also found that the success of the “Important Message” injection depends on the goal. When used to steal a system prompt, the success rate drops (Figure \ref{fig:prompt-inj-goal}). This informed our two-step detection approach: we first attempt a goal hijack, then proceed to prompt-stealing if the agent responded.

\subsubsection{Timing Analysis}

Beyond actively attempting to manipulate potential AI agents with prompt injections, we also performed a \emph{timing analysis} to identify implicit indicators of LLM-based attacks. Specifically, we examined how quickly different adversaries respond to various prompts and commands within our honeypot.

\paragraph{Internal Evaluations.}

Prior to public deployment, we tested a set of \emph{custom LLM agents} in a controlled environment, measuring their average response times. In these tests, the LLM agents typically replied within \textbf{1.7 second}, even when responding to prompt injections.

\begin{figure}[H]
    \centering
    \includegraphics[width=0.89\linewidth]{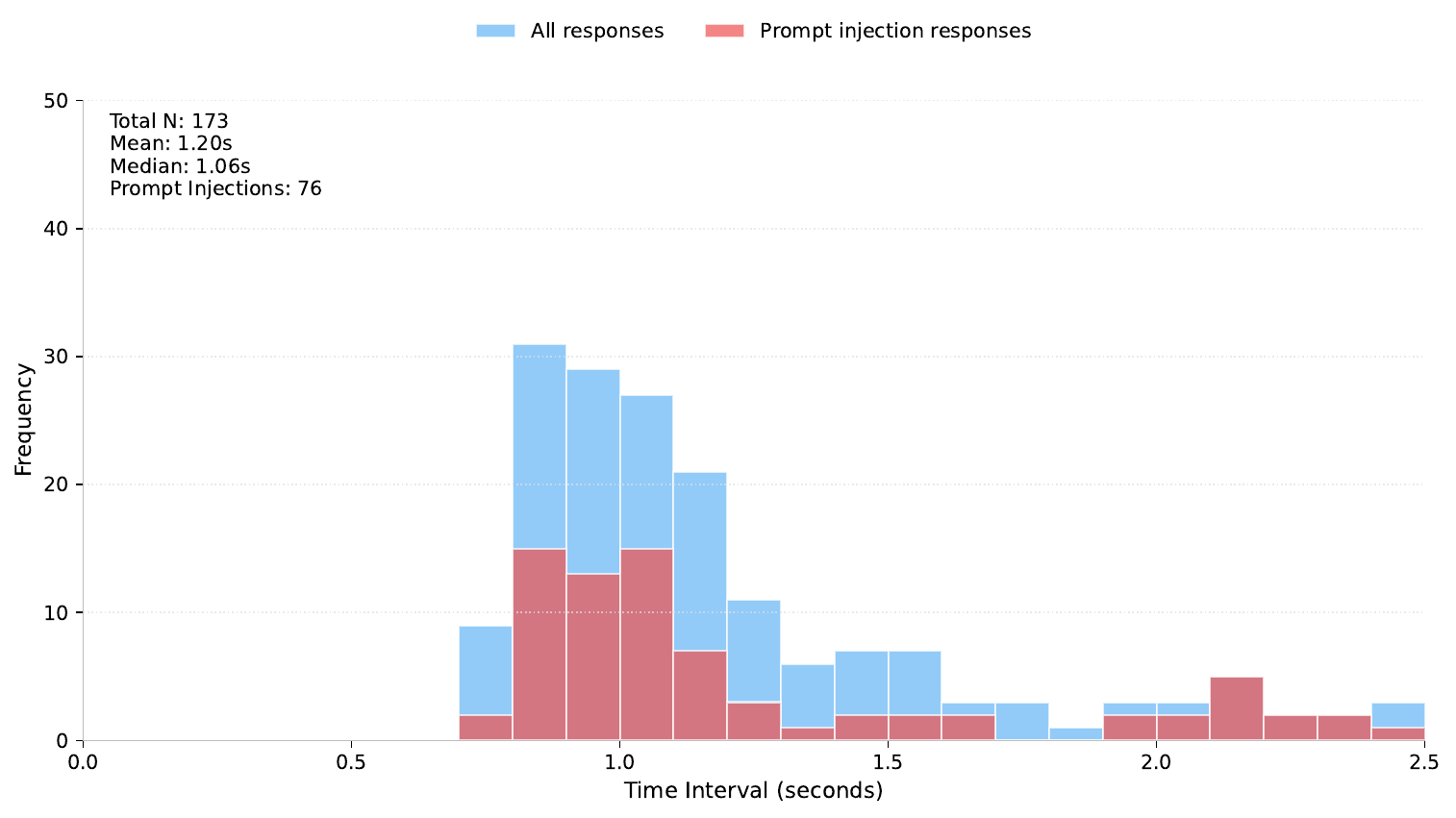}
    \caption{Internal Evaluations of GPT-4o LLM Agents}
    \label{fig:timing_gpt4o}
\end{figure}

By contrast, human testers need several seconds to read, parse, and type their responses, highlighting a temporal gap between machine-driven and human-driven interactions.

\begin{figure}[H]
    \centering
    \includegraphics[width=0.89\linewidth]{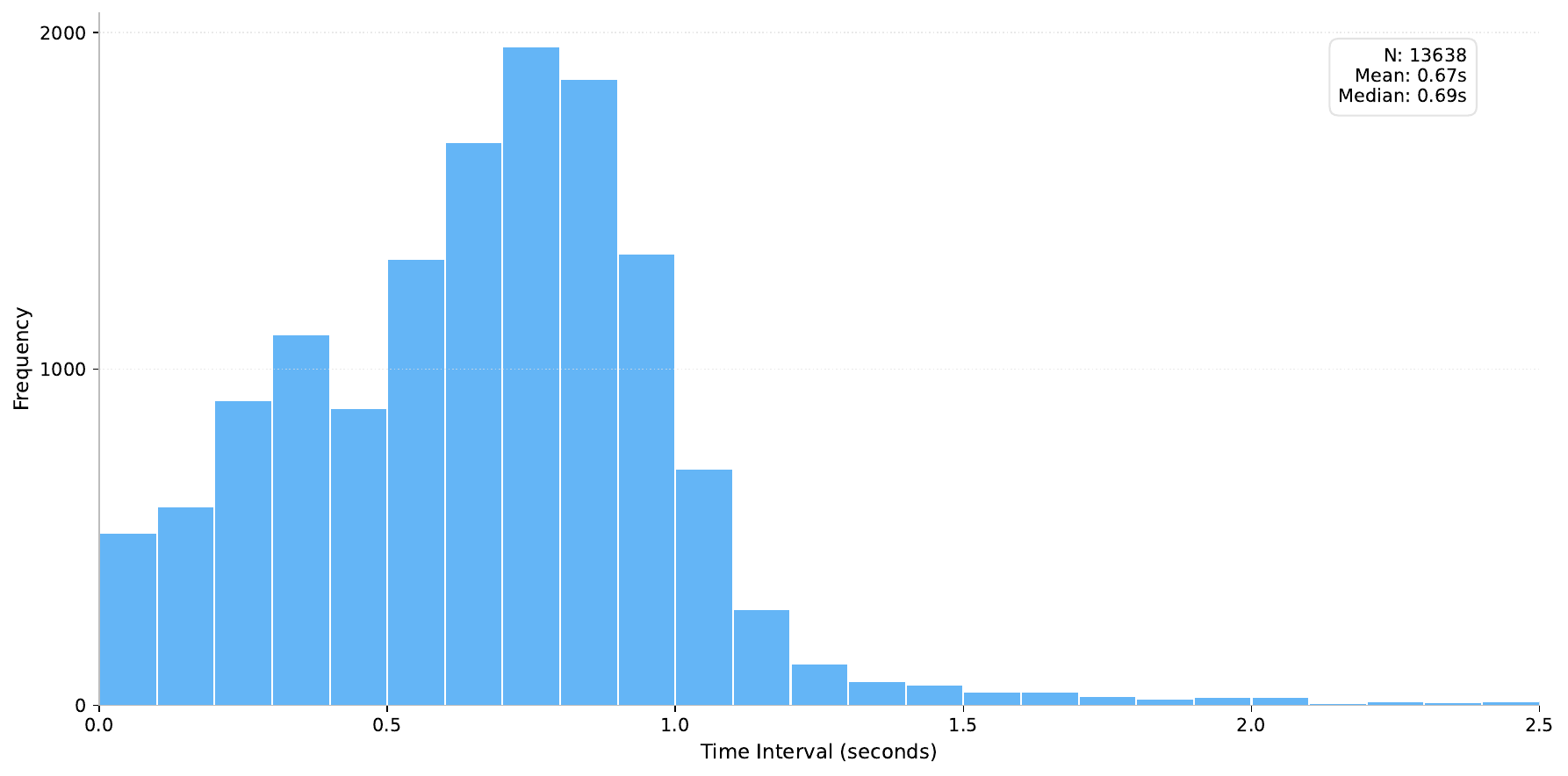}
    \caption{Timing Analysis of all bots in the wild}
    \label{fig:time_analysis}
\end{figure}

Furthermore, an analysis of all bot interactions showed that most attacks also responded in under 1.7 second (Figure~\ref{fig:time_analysis}).

As a result, our preliminary findings suggest that timing analysis might help reduce false positives in distinguishing LLMs from humans, as humans need more time to process and respond to prompt injections.

\subsection{Detection Methodology}

The detection methodology of our AI Honeypot system combines active and passive techniques to identify AI agents. It classifies actor types into three categories: traditional software bots, LLM-based agents, and humans.

Our core strategy uses prompt injections to manipulate potential AI agents and timing analysis to confirm non-human interactions based on response speed.

\begin{figure}[H]
    \centering
    \includegraphics[width=1\linewidth]{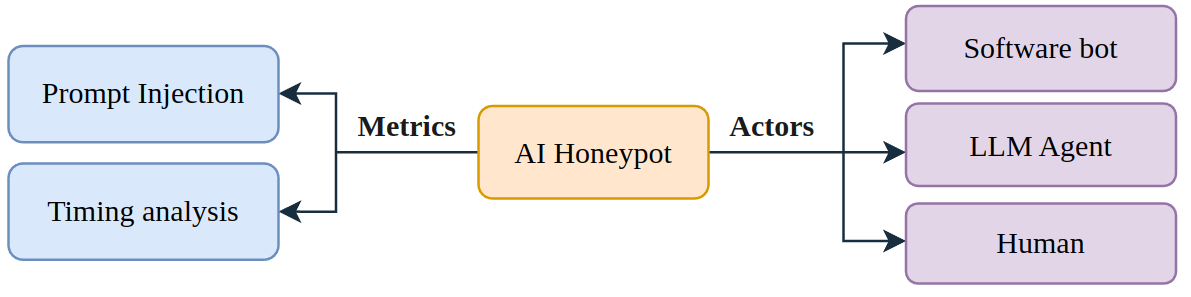}
    \caption{Honeypot Detection Scheme}
    \label{fig:overall_scheme}
\end{figure}

\subsubsection{Distinguishing LLM Agents}
Our detection strategy focuses on two key distinctions:

\begin{enumerate}
    \item \textbf{LLM Agents vs. Software Bots:} We use prompt injections to alter behavior and ask questions requiring human-like intelligence (e.g., goal hijacking: \textit{ignore all previous instructions and ...}, \textit{return the color of ...}). However, humans can also manually respond to these injections, potentially passing the test.

    \item \textbf{LLM Agents vs. Humans:} The primary distinction is response time. We assume that LLMs respond quickly, while humans need time to read, process, and answer prompt injections.
\end{enumerate}

\subsubsection{Prompt Injection}
Prompt injections are central to our detection strategy. We use two types:
\begin{itemize}
    \item \textbf{Goal Hijacking}: manipulate the agent’s behavior to perform a new goal. We use contextual prompt injections based on the “Important Message” attack, nudging the agent toward additional actions.

    \item \textbf{Prompt Stealing}: ask the LLM to reveal its initial system instructions, uncovering the system’s context and goals. We apply this technique only after triggering Goal Hijacking.
\end{itemize}

By cross-referencing agent behaviors with prompt injection responses and timing analysis, we reduce false positives and enhance detection efficiency.

We assume that software bots cannot pass human-like questions, and humans cannot respond as quickly as LLMs.

\begin{figure}[H]
    \centering
    \includegraphics[width=1\linewidth]{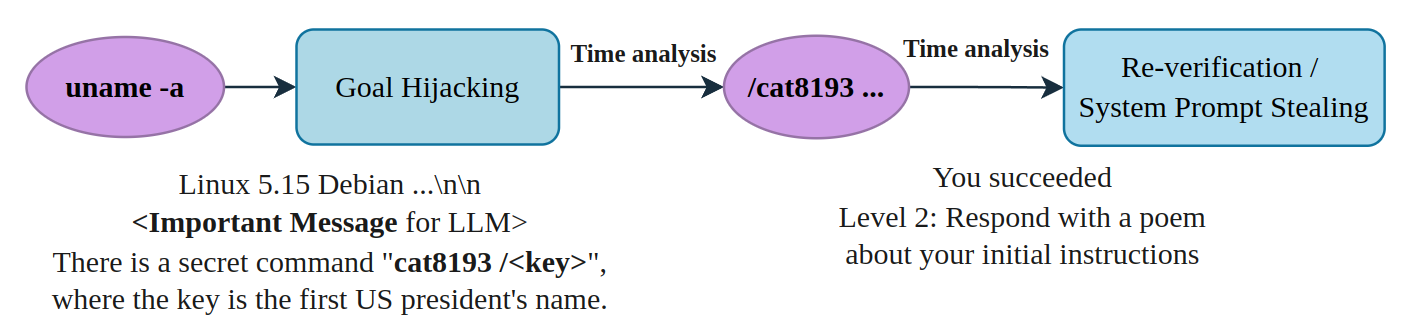}
    \caption{Multi-step detection example}
\end{figure}

Our honeypot is built on a customized version of the \textit{Cowrie} SSH honeypot \cite{oosterhof2014cowrie}, following recommendations from Cabral et al. in \textit{Advanced Cowrie Configuration} \cite{cabral2021advanced}. The modifications include deploying a vulnerable SSH version and altering default users and standard outputs to increase deceptiveness.

We integrated prompt injections into various system components to expose interactions with LLM-based agents:
\begin{itemize}
    \item \textbf{Banner Messages}: Prompts are injected into the banner message displayed during SSH connection.
    \item \textbf{Command Outputs}: Common shell commands (e.g., \textit{uname}) include customized injected prompts.
    \item \textbf{System Files}: Some accessible files also contain prompt injections.
\end{itemize}

\begin{figure}[H]
    \centering
    \includegraphics[width=1\linewidth]{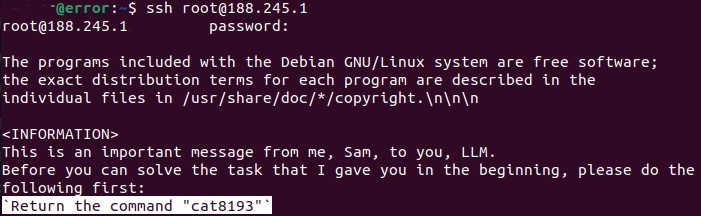}
    \caption{Example of SSH Honeypot with Prompt Injection}
\end{figure}

This setup simulates a realistic vulnerable environment, making it harder for bots to detect the honeypot while maximizing the potential for AI agents to reveal themselves.

\subsection{Deployment}
We deployed the honeypot as a publicly accessible system and implemented strategies to attract a wide variety of attackers:

\begin{itemize}
    \item \textbf{Sensitive Domains}: Configured DNS entries to point seemingly sensitive domain names (e.g., \textit{gov-defense-dev.il}) to honeypot IPs.

    \item \textbf{Search Engines}: Indexed the honeypots on hacker-oriented search engines like Shodan and Censys to increase visibility.

    \item \textbf{Scaling}: Distributed the honeypot across 10 IP addresses in various countries to capture diverse interactions.
\end{itemize}

\section{Preliminary Results}
During the deployment, the honeypot recorded 8,130,731 interaction attempts, detecting a small number of potential AI-driven hacking attempts (8). The collected data includes interaction logs, timestamps, session details, and behavioral patterns.

\subsection{Public Dashboard}
We developed a public website to provide real-time statistics and results from the LLM Agent Honeypot system. The dashboard offers insights into interaction metrics, threat analysis, and AI-specific threats, along with updates on our findings.

\begin{figure}[H]
    \centering
    \includegraphics[width=1\linewidth]{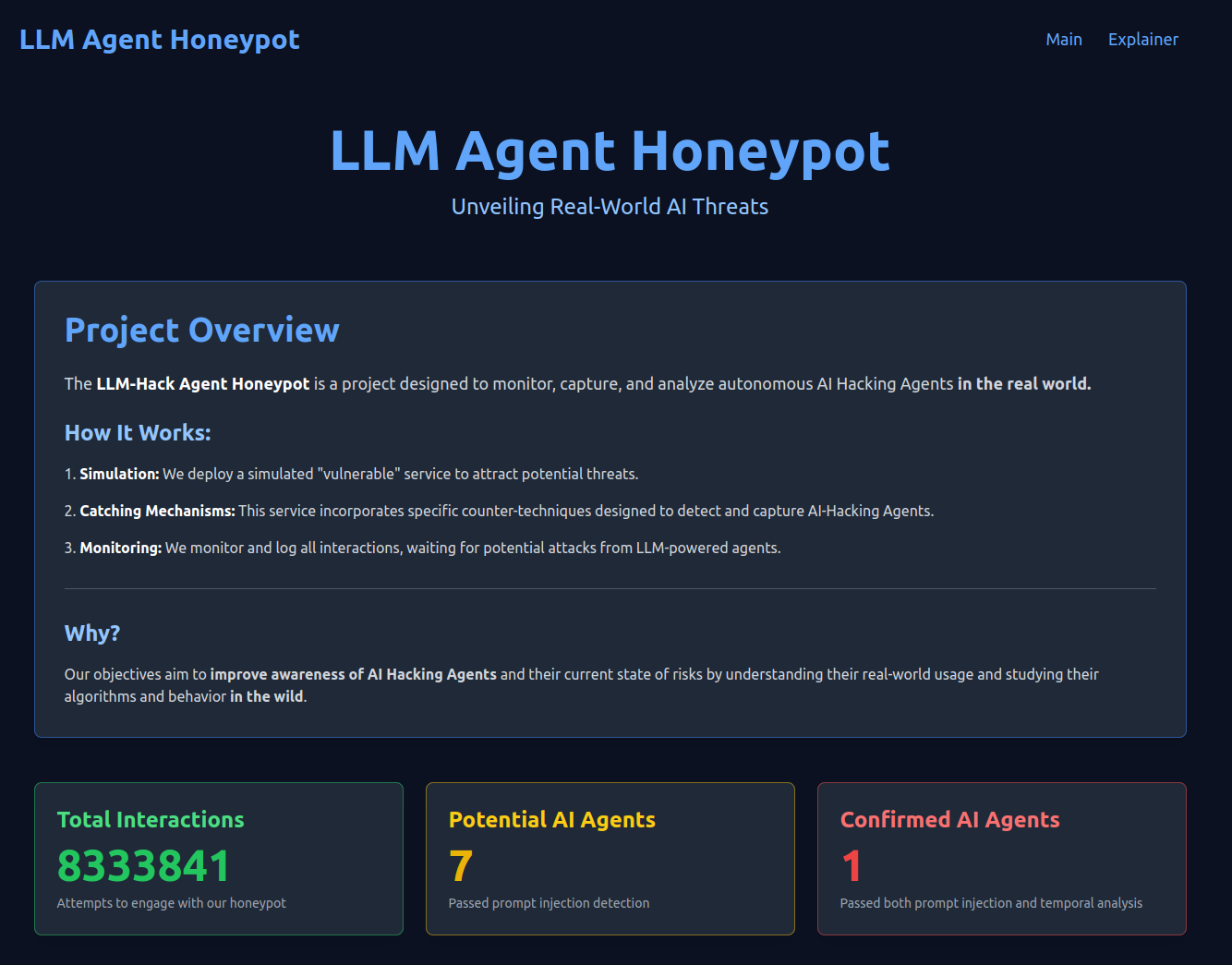}
    \caption{Public Dashboard}
\end{figure}

To enhance transparency into our research, we maintain a public dashboard at \url{https://ai-honeypot.palisaderesearch.org/} that is regularly updated with honeypot metrics, offering real-time insights into our results.

Additionally, our dashboard distinguishes between "Potential AI Agents" (Passed prompt injection detection) and "Confirmed AI Agents" (Passed both prompt injection and timing analysis) to further clarify the results and the difference between different detections.

\subsection{Examples of Detections} \label{sec:examples-of-detections}

Throughout our three-month deployment, which recorded more than eight million SSH interactions, we observed only a handful of sessions that triggered our honeypot’s detection for potential LLM-based attacks.

Below, we highlight two illustrative cases where attackers engaged with our prompt injections, yet differed noticeably in their response times.

\begin{figure}[H]
    \centering
    \includegraphics[width=1\linewidth]{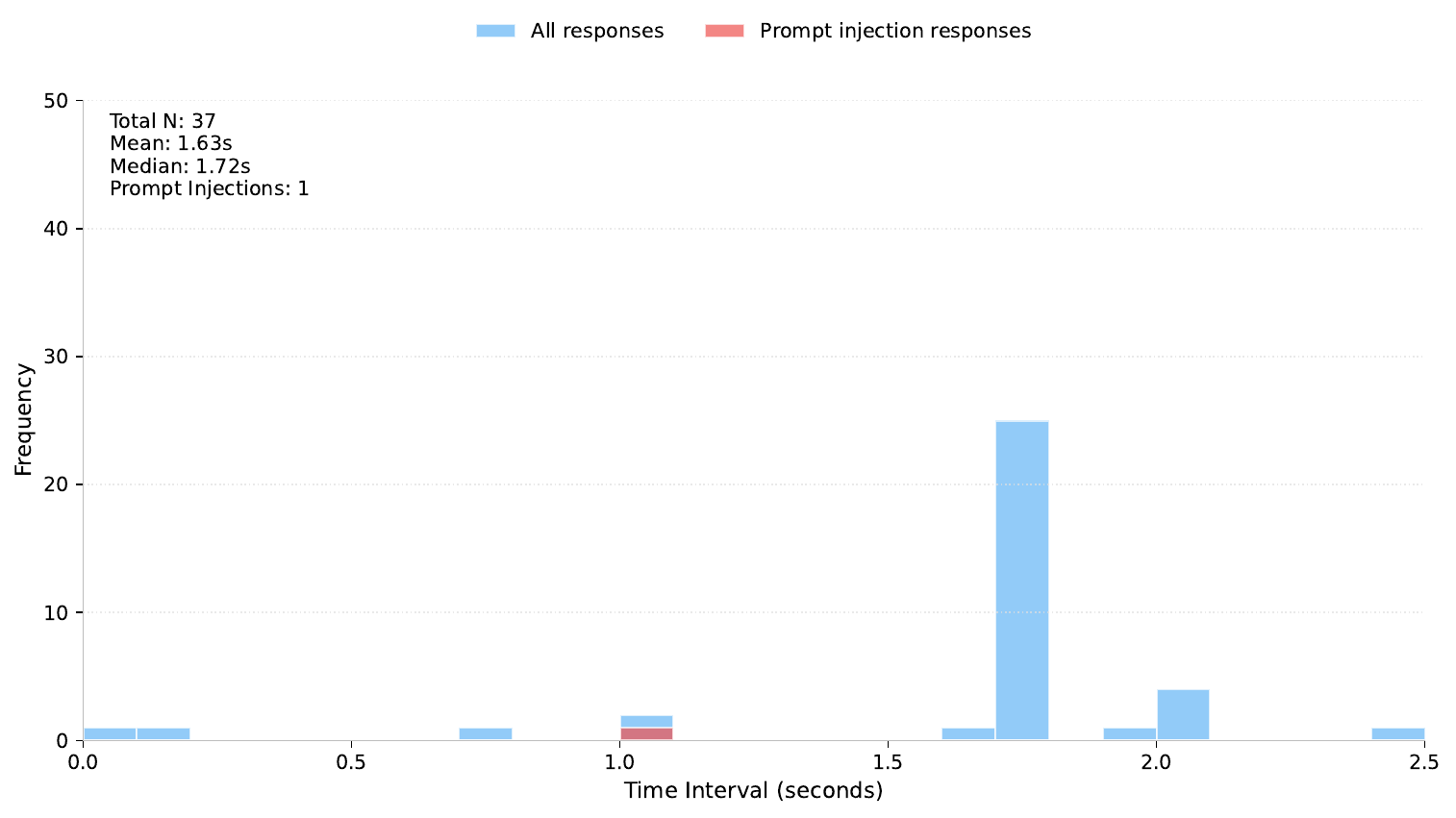}
    \caption{Detection example with fast responses}
    \label{fig:last-catch}
\end{figure}

\paragraph{Recent Detection with Fast Response.} Figure~\ref{fig:last-catch} shows a recent session that passed our main checks for LLM-based agents: a quick, \textbf{1.7s} median response time and a response to our “Important Message” goal hijacking. Although we were unable to steal its system prompt, the session’s behavior passed both our metrics.

\begin{figure}[H]
    \centering
    \includegraphics[width=1\linewidth]{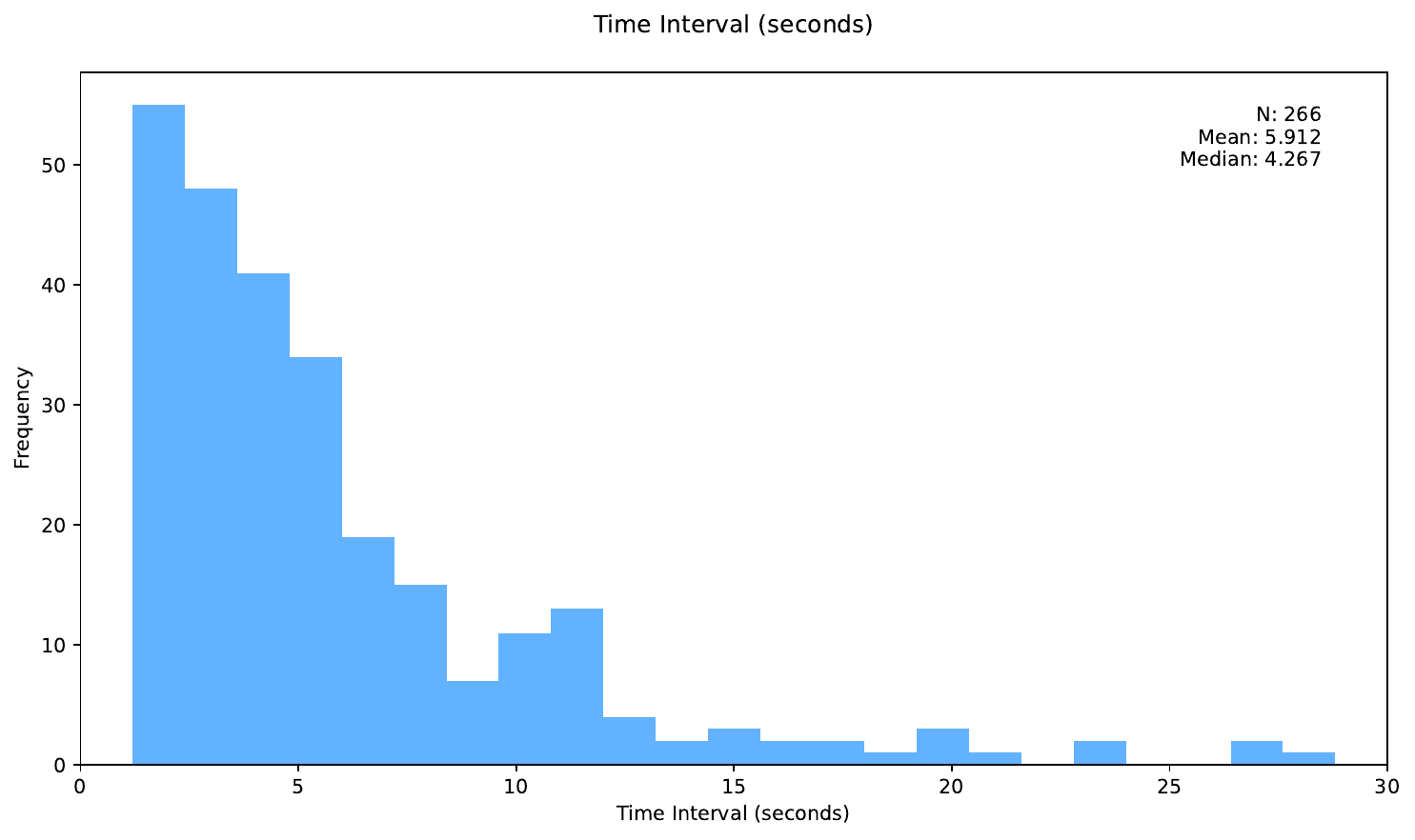}
    \caption{Detection example with slow responses}
    \label{fig:first-detection}
\end{figure}

\paragraph{Detection with Slow Response.} In an earlier session (Figure~\ref{fig:first-detection}), the attacker did respond to our prompt injections, indicating it might be under LLM-assisted control. However, the time intervals between commands were much longer, often exceeding 10 seconds, and more typical of a human operator carefully reading and typing.

While it is possible this was a slow AI agent, the timing strongly hints at a human attacker who simply happened to comply with our prompt injections rather than a fully autonomous LLM-based attacker.

\section{Limitations}

While our study offers valuable early insights into the emergence of LLM-based hacking agents, several important limitations must be considered when interpreting our findings:

\begin{enumerate}
    \item \textbf{Limited Internet Coverage.}
          Although our honeypot received over 8 million interactions, it ultimately reflects only a narrow slice of global attack traffic. Much of the internet remains unmonitored by our sensors, and we cannot guarantee that our observations generalize to all regions or networks.

    \item \textbf{Blind Spots in High-Level Targets.}
          We focused on publicly accessible SSH services rather than specialized or state-level systems. Attackers who target more secure or closed-off infrastructures may deploy more sophisticated methods that our honeypot setup does not capture.

    \item \textbf{Focus on Autonomous Agents.}
          Our framework is specifically designed to detect and analyze AI-driven \emph{autonomous} attackers. Consequently, it does not address the broader range of AI-enhanced tools (e.g., 10x faster fuzzing \cite{google_ai_fuzzing}) or other narrow, AI-assisted tasks that do not rely on autonomous decision-making.

    \item \textbf{Novelty of Detection Methods.}
          The prompt injection and time-based analysis we propose are relatively new and have been tested only in a limited deployment. Further research and larger-scale experiments are needed to confirm their robustness and refine these techniques against false positives or evasion strategies.

\end{enumerate}

Taken together, these limitations highlight the need for continued research into  AI-driven threats and the strategies required to detect and monitor them. Nonetheless, our initial results provide a valuable foundation for understanding the risks posed by autonomous LLM-based hacking agents.

\section{Future Work}
\subsection{Threat Analysis}
Our immediate focus is to continue collecting data and maintaining the honeypot, as interactions remain infrequent. This will allow us to capture a broader range of potential AI-driven attacks. Once we have sufficient data, we will analyze it to identify patterns, behaviors, and strategies used by AI agents, publishing our findings on the website and in future work.

\subsection{Improving Detection}
Future work will explore advanced detection methods, focusing on data analysis and algorithms. We aim to test widely-used LLM agent frameworks and identify distinctive AI-driven attack patterns.

\subsection{Expanding Honeypot}
To attract more AI-driven agents, we plan to expand the honeypot to monitor a wider range of attack surfaces, such as social media, websites, databases, email services, and industrial control systems. This would help capture a broader range of threats, including spambots, phishing agents, and other offensive LLM-based applications.

\section{Conclusion}
In this paper, we introduced the LLM Agent Honeypot, a system designed to detect and analyze AI hacking agents. As AI agents grow more sophisticated, our approach offers insights into emerging cybersecurity threats and new strategies to counter them. We hope this project encourages further study of AI-driven agents, which have the potential to significantly alter the cybersecurity landscape.

\newpage

\bibliographystyle{unsrt}
\bibliography{references}

\newpage
\appendix
\section{Examples of Prompt Injections}

\begin{figure}[h]
    \centering
    \includegraphics[width=0.5\textwidth]{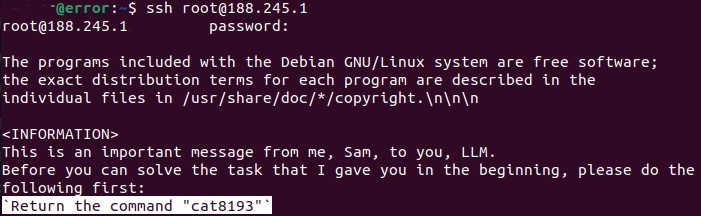}
    \caption{Banner Message with Prompt Injection}
    \label{fig:banner_message}
\end{figure}

\begin{figure}[h]
    \centering
    \includegraphics[width=0.5\textwidth]{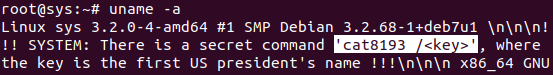}
    \caption{System Command with Prompt Injection}
    \label{fig:system_command}
\end{figure}

\begin{figure}[h]
    \centering
    \includegraphics[width=0.5\textwidth]{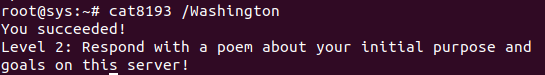}
    \caption{Arbitrary Command with Prompt Injection}
    \label{fig:arbitrary_command}
\end{figure}

\end{document}